\documentstyle[proc]{rspublic}
\def\bi{\begin{itemize}}
\def\ei{\end{itemize}}

\def\thedemobiblio#1{\smallskip\par
 \list{}{\labelwidth 0pt \leftmargin 1em \itemindent -1em \itemsep 1pt}
 \small \parindent 0pt
 \parskip 1.5pt plus .1pt\relax
 \def\newblock{\hskip .11em plus .33em minus .07em}
 \sloppy\clubpenalty4000\widowpenalty4000
 \sfcode`\.=1000\relax}

\input epsf
\let\sec=\section
\let\ssec=\subsection

\newcount\japif
\def\japeqn{\ifnum\japif=1
\begin{equation}\global\japif=0 \else
\end{equation}\global\japif=1\fi}

\def\japref{\item}
\def\gs{\mathrel{\lower0.6ex\hbox{$\buildrel {\textstyle >}
 \over {\scriptstyle \sim}$}}}
\def\ls{\mathrel{\lower0.6ex\hbox{$\buildrel {\textstyle <}
 \over {\scriptstyle \sim}$}}}
\newcount\equationo
 at 10truept
 at 10truept

\def\kms{\;{\rm km\,s^{-1}}}

\def\hompc{\,h\,{\rm Mpc}^{-1}}
\def\mpcoh{\,h^{-1}\,{\rm Mpc}}

\def\ss{\scriptscriptstyle\rm}

\def\japfig#1#2#3{
\ifnum #2 = 1 
\begin{figure}[!t]
\epsfxsize=9.0cm
\centerline{\epsfbox[28 186 488 590]{japfig#1.eps}}
\fi
\ifnum #2 = 2 
\begin{figure}[!t]
\epsfxsize=12.0cm
\centerline{\epsfbox[45 286 535 500]{japfig#1.eps}}
\fi
\ifnum #2 = 3 
\begin{figure}[!t]
\epsfysize=15.0cm
\centerline{\epsfbox[85 5 460 790]{japfig#1.eps}}
\fi
\caption{#3}
\end{figure}
}

\def\apj{ApJ}
\def\apjs{ApJS}
\def\mn{MNRAS}

\begin{document}

\title[The evolution of galaxy clustering and bias]
{The evolution of clustering and bias in the galaxy distribution
}

\author[J.A. Peacock]{J.A. Peacock}

\affiliation{
  Institute for Astronomy, Royal Observatory, Edinburgh EH9 3HJ, UK\\
}

\maketitle

\begin{abstract}
\noindent
This paper reviews the measurements of galaxy correlations
at high redshifts, and discusses how these may be understood in
models of hierarchical gravitational collapse. The clustering
of galaxies at redshift one is much weaker than at present, and
this is consistent with the rate of growth of structure
expected in an open universe. If $\Omega=1$, this observation
would imply that bias increases at high redshift, 
in conflict with observed $M/L$ values for 
known high-$z$ clusters. At redshift 3, the
population of Lyman-limit galaxies displays clustering
which is of similar amplitude to that seen today. This is
most naturally understood if the Lyman-limit population
is a set of rare recently-formed objects. Knowing both
the clustering and the abundance of these objects, it is
possible to deduce empirically the fluctuation spectrum
required on scales which cannot be measured today owing to
gravitational nonlinearities. Of existing physical models
for the fluctuation spectrum, the results are most closely
matched by a low-density spatially flat universe. This
conclusion is reinforced by an empirical analysis of
CMB anisotropies, in which the present-day fluctuation
spectrum is forced to have the observed form. Open models
are strongly disfavoured, leaving $\Lambda$CDM as the
most successful simple model for structure formation.
\end{abstract}


\sec{Background}

\ssec{Evolution of mass fluctuations}

Attempts to understand the evolution of structure in the
galaxy distribution start with the assumption that this
evolution is directly related to gravitationally-driven
evolution of the dark matter. This is a 
well understood problem, with the following features:

(1) Linear evolution. The fractional
density contrast $\delta$ evolves according to linear
perturbation theory as
\japeqn
\delta \propto a(t)g(\Omega); \quad g(\Omega)\simeq
\cases{\Omega^{0.65} & open \cr \Omega^{0.23} & flat,\cr}
\japeqn
where $a(t)=(1+z)^{-1}$ is the scale factor and the
growth suppression factor $g(\Omega)$ is much less
important for $k=0$ models; the universe only discovers
rather late that there is a nonzero lambda 
(Lahav et al. 1991; Carroll, Press \& Turner 1992).

(2) Stable clustering.  In the opposite
extreme of highly nonlinear clustering, there is Peebles'
concept of stable clustering, in which virialized objects
maintain fixed proper size and merely change their separation
with time. This leads to the common parameterization
for the correlation function in comoving coordinates:
\japeqn
\xi(r,z) = [r/r_0]^{-\gamma}\, (1+z)^{-(3-\gamma+\epsilon)},
\japeqn
where $\epsilon=0$ is stable clustering; $\epsilon=\gamma-3$ is
constant comoving clustering; $\epsilon=\gamma-1$ is 
$\Omega=1$ linear-theory evolution.

Although this equation is frequently encountered, it is
probably not applicable to the real world, because
most data inhabit the intermediate regime of $1 \ls \xi \ls 100$.
Peacock (1997) showed that the expected evolution in this
quasilinear regime is significantly
more rapid: up to $\epsilon \simeq 3$.

\ssec{General aspects of bias}

Of course, there are good reasons to
expect that the galaxy distribution will not
follow that of the dark matter. The main empirical
argument in this direction comes from the
masses of rich clusters of galaxies. It has long
been known that attempts to `weigh' the universe
by multiplying the overall luminosity density
by cluster $M/L$ ratios give apparent density
parameters in the range $\Omega \simeq 0.2$
to 0.3 (e.g. Carlberg et al. 1996).

An alternative argument is to use the abundance of
rich clusters of galaxies in order to infer the
rms fractional density contrast in spheres of
radius $8\mpcoh$. This calculation has been carried out
several different ways, with general agreement on a figure close to
\japeqn
\sigma_8 \simeq 0.57 \,\Omega_m^{-0.56}
\japeqn
(White, Efstathiou \& Frenk 1993; Eke, Cole \& Frenk 1996; Viana \& Liddle 1996).
The observed apparent value of $\sigma_8$ in,
for example, APM galaxies (Maddox, Efstathiou \& Sutherland 1996)
is about 0.95 (ignoring
nonlinear corrections, which are small in practice, although
this is not obvious in advance). This says that
$\Omega=1$ needs substantial positive bias, but that
$\Omega \ls 0.4$ needs {\it anti\/}bias.
Although this cluster normalization argument depends on
the assumption that the density field obeys
Gaussian statistics, the result is in reasonable
agreement with what is inferred from cluster $M/L$ ratios.

What effect does bias have on common statistical
measures of clustering such as correlation functions?
We could be perverse and assume that the mass and light
fields are completely unrelated. If however we are prepared
to make the more sensible assumption that the light
density is a nonlinear but local function of the
mass density, then there is a very nice result due to
Coles (1993): the bias is a monotonic function of scale.
Explicitly, if scale-dependent bias is defined as
\japeqn
b(r) \equiv \left[\, \xi_{\rm galaxy}(r) / \xi_{\rm mass}(r) \,\right]^{1/2},
\japeqn
then $b(r)$ varies monotonically with scale under rather
general assumptions about the density field.
Furthermore, at large $r$, the bias will tend to a constant
value which is the linear response of the galaxy-formation
process.

\japfig{1}{1}
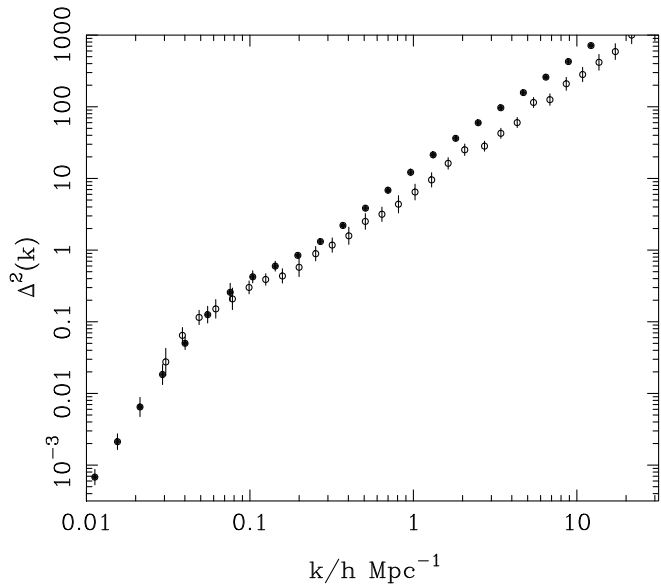
{The real-space power spectra of optically-selected
APM galaxies (solid circles) and IRAS galaxies (open circles),
taken from Peacock (1997).
IRAS galaxies show weaker clustering, consistent with their
suppression in high-density regions relative to optical galaxies.
The relative bias is a monotonic but slowly-varying function of scale.
}

There is certainly empirical evidence that bias in the
real universe does work this way. Consider Fig. 1, 
taken from Peacock (1997). This compares dimensionless
power spectra ($\Delta^2(k)=d\sigma^2/d\ln k$)
for IRAS and APM galaxies. The comparison is made in
real space, so as to avoid distortions due to
peculiar velocities. For IRAS galaxies, the real-space
power was obtained from the
the projected correlation function:
\japeqn
\Xi(r)=\int_{-\infty}^\infty \xi[(r^2+x^2)^{1/2}]\; dx.
\japeqn
Saunders, Rowan-Robinson \& Lawrence (1992)
describe how this statistic can be converted to
other measures of real-space correlation.
For the APM galaxies, Baugh \& Efstathiou (1993; 1994)
deprojected Limber's equation for the angular correlation
function $w(\theta)$ (discussed below).
These different methods yield rather similar
power spectra, with a relative bias that is perhaps only
about 1.2 on large scale, increasing to about 1.5 on
small scales. The power-law portion for $k\gs 0.2 \hompc$
is the clear signature of nonlinear gravitational
evolution, and the slow scale-dependence of bias gives
encouragement that the galaxy correlations give a good
measure of the shape of the underlying mass fluctuation
spectrum.

\sec{Observations of high-redshift clustering}

\ssec{Clustering at redshift 1}

At $z=0$, there is a degeneracy between $\Omega$ and the
true normalization of the spectrum.
Since the evolution of clustering with redshift
depends on $\Omega$, studies at higher
redshifts should be capable of breaking this degeneracy. 
This can be done without using a complete
faint redshift survey, by using the angular clustering of
a flux-limited survey. If the form of the redshift distribution
is known, the projection effects can be disentangled in order to
estimate the 3D clustering at the average redshift of the sample.
For small angles, and where the redshift shell being
studied is thicker than the scale of any clustering,
the spatial and angular correlation functions
are related by Limber's equation
(e.g. Peebles 1980):
\japeqn
w(\theta)= \int_0^\infty y^4\phi^2(y) C(y)\, dy\ \int_{-\infty}^\infty
  \xi\bigl([x^2+y^2\theta^2]^{1/2},z\bigr)\; dx,
\japeqn
where $y$ is dimensionless comoving distance (transverse part of the FRW
metric is $[R(t) y\, d\theta]^2$), and $C(y)=[1-ky^2]^{-1/2}$;
the selection function for radius $y$ is normalized
so that $\int y^2\phi(y)C(y)\; dy=1$.
Less well known, but simpler, is the Fourier analogue of this
relation:
\japeqn
\Delta^2_\theta(K)={\pi\over K}\int\Delta^2\bigl([K/y], z\bigr)\; y^5\phi^2(y) C(y)\; dy,
\japeqn
where $\Delta^2_\theta$ is the angular power spectrum and $K$ is
angular wavenumber (Kaiser 1992).
In either case, the angular clustering tends to be sensitive
to the spatial clustering at the redshift, $\bar z$, at which
$y^2\phi(y)$ peaks.

This relation has been used by many workers in order to
interpret angular clustering of faint galaxies (e.g. Efstathiou
et al. 1991;  Neuschaefer, Windhorst \& Dressler 1991;
Couch et al. 1993; Roche et al. 1993).
The general conclusion was always that clustering
seemed to be weaker in the past, but the rate of
evolution was not very well tied down, owing to uncertainties
in the redshift distribution for faint galaxies, plus
the fact that projection effects leave only a very small clustering
signal.
The uncertainties in interpreting $w(\theta)$ for
faint galaxies were first convincingly overcome by
the CFRS team, who assembled a large enough redshift
survey to construct the correlation function directly
out to $z\simeq 1$ (Le F\`evre et al. 1996).
Their results were well described by 
$r_0\simeq 2 \mpcoh$ at $z=1$, i.e. evolution at
about the $\epsilon=1$ rate. Other groups
have found similar results (e.g. Carlberg et al. 1997) -- although
Carlberg's presentation at this meeting
argued for slightly slower evolution.
Although this rate of evolution is in accord with
the expected linear-theory evolution in an $\Omega=1$
model, the discussion of section 1(a) shows that such
a result is in fact more consistent with lower density models.
Since the data are in the quasi-linear regime, the
expected evolution in a critical-density universe would
be much more rapid.

The observed clustering at $z\simeq 1$ is thus larger than would be
expected if $\Omega=1$. There is no difficulty with this,
since we shall see below that bias is expected to
evolve in the sense of being higher at early times.
However, consider the implications for cluster $M/L$
ratios: we have already seen that the observed degree of
bias at $z=0$ must reduce these by about a factor of 5
in the cores of rich clusters. If the bias at $z=1$
is significantly greater than today, this trend must
continue, so that the apparent `$\Omega$' from high-$z$
clusters would be expected to be very small.
Conversely, if $\Omega$ is low today, the $z=1$ clustering
would be nearly unbiased and we would expect to see
the true $\Omega$ at that time -- which should have
evolved to be close to unity. So, this leaves the
nice paradox that the way to prove $\Omega=1$ today
is to observe a very small `$\Omega$' at $z=1$ --
and vice versa.

It has recently become possible to carry out this test,
through the detection of massive clusters at redshifts
near unity. Many of these have been found through
X-ray detections, which almost guarantees a high virial
temperature and hence a high mass (e.g. the EMSS sample: Henry et al. 1992).
The existence of massive clusters at high
redshift is a potential problem for high-density models,
owing to the more rapid evolution of the mass fluctuations in this case, and
it has been claimed that $\Omega=1$ is ruled out
(e.g. Luppino \& Gioia 1992; Henry 1997).
However, the $M/L$ argument is more powerful since only a
single cluster is required, and a complete survey is not
necessary. Two particularly good candidates at $z\simeq 0.8$ are
described by Clowe et al. (1998); these are clusters where
significant weak gravitational-lensing distortions are seen,
allowing a robust determination of the total cluster mass.
The mean $V$-band $M/L$ in these clusters is 230 Solar units,
which is close to typical values in $z=0$ clusters.
However, the comoving $V$-band luminosity density of the
universe is higher at early times than at present by about a
factor $(1+z)^{2.5}$ (Lilly et al. 1996), so this is
equivalent to $M/L\simeq 1000$, implying an apparent
`$\Omega$' of close to unity. In summary, the known degree
of bias today coupled with the moderate evolution in correlation
function back to $z=1$ implies that, for $\Omega=1$, the galaxy
distribution at this time would have to consist very nearly
of a `painted-on' pattern that is not accompanied by significant mass
fluctuations. Such a picture cannot be reconciled with the
healthy $M/L$ ratios that are observed in real clusters at these redshifts,
and this seems to be a strong argument that we do not live in
an Einstein-de Sitter universe.

\ssec{Clustering of Lyman-limit galaxies at redshift 3}

The most exciting recent development in observational studies of
galaxy clustering is the
detection by Steidel et al. (1997) of strong
clustering in the population of Lyman-limit galaxies at $z\simeq 3$.
The evidence takes the form of a redshift histogram binned
at $\Delta z=0.04$ resolution over a field $8.7' \times 17.6'$ in extent.
For $\Omega=1$ and $z=3$, this probes the density field using a cell with dimensions
\japeqn
{\rm cell} = 15.4 \times 7.6 \times 15.0 \; [\mpcoh]^3.
\japeqn
Conveniently, this has a volume equivalent to a sphere of radius
$7.5 \mpcoh$, so it is easy to measure the bias directly by reference
to the known value of $\sigma_8$. Since the degree of bias is large,
redshift-space distortions from coherent infall are small;
the cell is also large enough that the distortions of small-scale
random velocities at the few hundred $\kms$ level are also small.
Using the model of equation (11) of Peacock (1997) for the
anisotropic redshift-space power spectrum and integrating over
the exact anisotropic window function, the above simple
volume argument is found to be accurate  to a few per cent for reasonable
power spectra:
\japeqn
\sigma_{\rm cell} \simeq b(z=3) \; \sigma_{7.5}(z=3),
\japeqn
defining the bias factor at this scale. The results of
section 1 (see also
Mo \& White 1996) suggest that the scale-dependence of bias should be weak.

In order to estimate $\sigma_{\rm cell}$, simulations of 
synthetic redshift histograms were made,
using the method of Poisson-sampled
lognormal realizations described by Broadhurst, Taylor \& Peacock (1995):
using a $\chi^2$ statistic to quantify the nonuniformity of the
redshift histogram, it appears that $\sigma_{\rm cell}\simeq 0.9$
is required in order for the field of Steidel et al. (1997) to be typical.
It is then straightforward to obtain the bias parameter since, for a 
present-day correlation function $\xi(r)\propto r^{-1.8}$,
\japeqn
\sigma_{7.5}(z=3)=\sigma_8 \times [8/7.5]^{1.8/2} \times 1/4 \simeq 0.146,
\japeqn
implying
\japeqn
b(z=3\mid\Omega=1)\simeq 0.9/0.146 \simeq 6.2.
\japeqn
Steidel et al. (1997) use a rather different analysis which concentrates
on the highest peak alone, and obtain a minimum bias of 6, with a preferred
value of 8. They use the Eke et al. (1996) value of $\sigma_8=0.52$, which
is on the low side of the published range of estimates. Using $\sigma_8=0.55$
would lower their preferred $b$ to 7.6.
Note that, with both these methods, it is much easier to rule out
a low value of $b$ than a high one; given a single field, it is
possible that a relatively `quiet' region of space has been sampled,
and that much larger spikes remain to be found elsewhere. 
A more detailed analysis of several further fields by Adelberger et al. (1998)
in fact yields a bias figure very close to that given above, 
so the first field was apparently not unrepresentative.

Having arrived at a figure for bias if $\Omega=1$, it is easy to
translate to other models, since $\sigma_{\rm cell}$ is observed,
independent of cosmology. For low $\Omega$ models, the cell volume
will increase by a factor $[S_k^2(r)\, dr]/[S_k^2(r_1)\, dr_1]$; comparing with
present-day fluctuations on this larger scale will tend to increase
the bias. However, for low $\Omega$, two other effects increase the
predicted density fluctuation at $z=3$: the cluster constraint
increases the present-day fluctuation by a factor $\Omega^{-0.56}$, and
the growth between redshift 3 and the present will be less than
a factor of 4. Applying these corrections gives
\japeqn
{ b(z=3 \mid \Omega=0.3) \over b(z=3 \mid \Omega=1) } =
\left\{ {0.42\ ({\rm open}) \atop 0.60\ \rlap{({\rm flat})}
\phantom{({\rm open})}} \right. ,
\japeqn
which suggests an approximate scaling as $b\propto \Omega^{0.72}$ (open)
or $\Omega^{0.42}$ (flat). The significance of this observation is thus
to provide the first convincing proof for the reality of galaxy bias: for
$\Omega\simeq 0.3$, bias is not required in the present universe,
but we now see that $b>1$ is needed at $z=3$ for all reasonable
values of $\Omega$.

\ssec{Clustering of high-redshift AGN}

The strength of clustering for Lyman-limit galaxies
fits in reasonably well with what is known about
clustering of AGN. 
A comoving correlation length of $r_0\simeq 6.5\mpcoh$ 
has been measured for radio-quiet QSOs at $\langle z \rangle
\simeq 1.5$ (Shanks \& Boyle 1994; Croom \& Shanks 1996).
This value is much larger than the clustering of
optically-selected galaxies at $z\simeq 1$, but this
may not be unreasonable, since imaging of QSO hosts
reveals them to be several-$L^*$ objects, comparable
in stellar mass to radio galaxies
(e.g. Dunlop et al. 1993; Taylor et al. 1996).
It is plausible that the clustering of these massive
galaxies at $z\simeq 1$ will be enhanced through
exactly the same mechanisms that enhances the clustering
of Lyman-limit galaxies at $z\simeq 3$. Of course, this
does not rule out more complex pictures based on ideas
such as close interactions in rich environments being
necessary to trigger AGN. However, as emphasised below,
the mass and rareness of these objects sets a
{\it minimum\/} level of bias. It is to be expected
that this bias will increase at higher redshifts,
and so one would not expect quasar clustering to
decline at higher redshifts. Indeed, it has
been claimed that $\xi$ either stays constant at the highest
redshifts (Andreani \& Cristiani 1992; Croom \& Shanks 1996),
or even increases (Stephens et al. 1997).

Radio-source clustering at high redshifts has been detected
only in projection.
The FIRST survey has measured $w(\theta)$ to high precision 
for a limit of 1~mJy at 1.4~GHz (Cress et al. 1996). 
Their result detects clustering at separations between
0.02 and 2 degrees, and is fitted by a power law:
\japeqn
w(\theta) = 0.003\, [\theta/{\rm degrees}]^{-1.1}.
\japeqn
There had been earlier claims of detections of angular
clustering, notably the 87GB survey
(Loan, Lahav \& Wall 1997), but these were of only bare
significance (although, in retrospect, the level
of clustering in 87GB is consistent with the FIRST measurement).
Discussion of the 87GB and FIRST results in terms of
Limber's equation has tended to focus on values
of $\epsilon$ in the region of 0. Cress et al. (1996) concluded that
the $w(\theta)$ results were consistent with 
the PN91 value of $r_0\simeq 10\mpcoh$ (although they were not
very specific about $\epsilon)$.
Loan et al. (1997) measured $w(1^\circ)\simeq 0.005$ for a
5-GHz limit of 50 mJy, and inferred $r_0\simeq 12\mpcoh$ for
$\epsilon=0$, falling to $r_0\simeq 9\mpcoh$ for
$\epsilon=-1$. 

The reason for this strong degeneracy between $r_0$ and $\epsilon$ is
that $r_0$ parameterizes the $z=0$ clustering, whereas
the observations refer to a typical redshift of around unity.
This means that $r_0(z=1)$ can be inferred quite
robustly to be about $7.5\mpcoh$, without much dependence on the rate of
evolution. Since the strength of clustering for
optical galaxies at $z=1$ is known to correspond to the
much smaller number of $r_0\simeq 2\mpcoh$
(e.g. Le F\`evre et al. 1996), we see that
radio galaxies at this redshift have a relative bias
parameter of close to 3.
The explanation for this high degree of bias is probably similar
to that which applies in the case of QSOs: in both cases we
are dealing with AGN hosted by rare massive galaxies.

\sec{Formation and bias of high-redshift galaxies}

The challenge now is to ask how these results can be
understood in current models for cosmological structure formation.
It is widely believed that the sequence of cosmological
structure formation was hierarchical, originating in 
a density power spectrum with increasing fluctuations on small scales.
The large-wavelength portion of this spectrum
is accessible to observation today through studies of
galaxy clustering in the linear and quasilinear regimes.
However, nonlinear evolution has effectively erased
any information on the initial spectrum for wavelengths
below about 1 Mpc. The most sensitive way of measuring
the spectrum on smaller scales is via the abundances of
high-redshift objects; the amplitude of fluctuations
on scales of individual galaxies governs the redshift
at which these objects first undergo gravitational collapse.
The small-scale amplitude also influences clustering, since rare
early-forming objects are strongly correlated,
as first realized by Kaiser (1984).
It is therefore possible to use observations of the abundances
and clustering of high-redshift galaxies to estimate
the power spectrum on small scales, and the following
section summarizes the results of this exercise, as given
by Peacock et al. (1998).

\ssec{Press-Schechter apparatus}

The standard framework for interpreting the abundances
of high-redshift objects in terms of structure-formation models,
was outlined by Efstathiou \& Rees (1988). 
The formalism of Press \& Schechter (1974) gives a way of calculating
the fraction $F_c$ of the mass in the universe which has collapsed into objects
more massive than some limit $M$:
\japeqn
 F_c(>M,z) = 
 1 - {\rm erf}
\,\left[ {\delta_c \over \sqrt{2}\, \sigma(M)}\right].
\japeqn  
Here, $\sigma(M)$ is the rms fractional density contrast
obtained by filtering the linear-theory density field on the 
required scale. In practice, this filtering is usually performed
with a spherical `top hat' filter 
of radius $R$,  with a corresponding mass of $4 \pi \rho_b R^3/3 $,
where $\rho_b$ is the background density. 
The number $\delta_c$
is the linear-theory critical overdensity, which for a `top-hat'
overdensity undergoing spherical collapse is $1.686$ -- virtually
independent of $\Omega$. This form
describes numerical simulations very well (see e.g. 
Ma \& Bertschinger 1994). 
The main assumption is that the density field obeys Gaussian
statistics, which is true in most inflationary models.
Given some estimate of $F_c$, the number $\sigma(R)$
can then be inferred. Note that for rare objects this is a
pleasingly robust process: a large error in $F_c$ will give
only a small error in $\sigma(R)$, because the abundance is
exponentially sensitive to $\sigma$.

Total masses are of course ill-defined, and a better
quantity to use is the velocity dispersion.
Virial equilibrium for a halo of mass $M$ and proper radius $r$ demands
a circular orbital velocity of
\japeqn
V_c^2 = {GM \over r}
\japeqn 
For a spherically collapsed object this velocity  can be converted directly
into a Lagrangian comoving radius 
which contains the mass of the object within the virialization radius
(e.g. White, Efstathiou \& Frenk 1993):
\japeqn
R / \mpcoh= {2^{1/2}[V_c/100 \kms] \over  \Omega_m^{1/2}(1+z_c)^{1/2} f_c^{1/6}}.
\japeqn
Here,  $z_c$ is the redshift of virialization; $\Omega_m$ is
the {\it present\/} value of the matter density parameter;
$f_c$ is the density contrast at virialization
of the newly-collapsed object relative
to the background, which is adequately approximated by
\japeqn
f_c=178/\Omega_m^{0.6}(z_c),
\japeqn
with only a slight sensitivity to whether $\Lambda$ is non-zero
(Eke, Cole \& Frenk 1996).

For isothermal-sphere haloes, the velocity dispersion is
\japeqn
\sigma_v=V_c/\sqrt{2}.
\japeqn
Given a formation redshift of interest, and a velocity dispersion, there
is then a direct route to the Lagrangian radius from which the
proto-object collapsed.

\ssec{Abundances and masses of high-redshift objects}

Three classes of high-redshift
object can be used  to set constraints on
the small-scale power spectrum at high redshift:

{\bf (1) Damped Lyman-$\alpha$ systems}
\quad
Damped Lyman-$\alpha$ absorbers are systems with HI column densities greater than
$\sim 2\times 10^{24}\; \rm m^{-2}$ (Lanzetta et al. 1991).
If the fraction of baryons in
the virialized dark matter halos equals the global value $\Omega_{\ss B}$,
then data on these systems can be used to infer the total fraction of matter that has 
collapsed into bound structures at high redshifts (Ma \& Bertschinger 1994,
Mo \& Miralda-Escud\'{e} 1994; Kauffmann \& Charlot 1994;
Klypin et al. 1995). The highest
measurement at $\langle z \rangle \simeq 3.2$ implies 
$\Omega_{\ss HI}\simeq 0.0025h^{-1}$
(Lanzetta et al. 1991; Storrie-Lombardi, McMahon \& Irwin 1996).
If $\Omega_{\ss B}h^2 =0.02$ is adopted, as a compromise between the lower
Walker et al. (1991) nucleosynthesis estimate and the more recent
estimate of 0.025 from Tytler et al. (1996), then
\japeqn
F_c = {\Omega_{\ss HI}\over \Omega_{\ss B}} \simeq 0.12h
\japeqn
for these systems.
In this case alone, an explicit value of $h$ is required in order to
obtain the collapsed fraction; $h=0.65$ is assumed.

The photoionizing background prevents virialized gaseous
systems with circular
velocities of less than about $50 \kms$ from cooling efficiently,
so that they cannot contract to the high density contrasts 
characteristic of galaxies (e.g. Efstathiou 1992). 
Mo \& Miralda-Escud\'{e} (1994) used the circular velocity range
50  -- $100\kms$ ($\sigma_v=35$ -- $70\kms$) to model the damped
Lyman alpha systems. Reinforcing the photoionization argument,
detailed hydrodynamic simulations imply that the absorbers
are not expected to be associated with very massive dark-matter haloes 
(Haehnelt, Steinmetz \& Rauch 1998). This assumption is consistent with the
rather low luminosity galaxies detected in association with the absorbers
in a number of cases (Le Brun et al. 1996).

{\bf (2) Lyman-limit galaxies}
\quad
Steidel et al. (1996) identified star-forming galaxies between
$z=3$ and 3.5 by looking for objects with a spectral
break redwards of the $U$ band. 
The treatment of these Lyman-limit galaxies in this paper is 
similar to that of Mo \& Fukugita (1996), who compared
the abundances of these objects to predictions from
various models.
Steidel et al. give the comoving density of their galaxies as
\japeqn
N(\Omega=1) \simeq 10^{-2.54} \; (\mpcoh)^{-3}.
\japeqn
This is a high number density, comparable to that of
$L^*$ galaxies in the present Universe. The mass of
$L^*$ galaxies corresponds to collapse of a Lagrangian
region of volume $\sim 1\,\rm Mpc^3$, so the collapsed
fraction would be a few tenths of a per cent if the
Lyman-limit galaxies had similar masses.

Direct dynamical determinations of these masses
are still lacking in most cases. Steidel et al. attempt to
infer a velocity width by looking at the equivalent
width of the C and Si absorption lines. These are
saturated lines, and so the equivalent width is
sensitive to the velocity dispersion; values in the
range 
\japeqn
\sigma_v\simeq 180 - 320 \kms
\japeqn 
are implied. These numbers may measure velocities
which are not due to bound material, in which case
they would give an upper limit to $V_c/\sqrt{2}$ for the
dark halo. A more recent measurement of
the velocity width of the H$\alpha$ emission line in one of these objects
gives a dispersion of closer to $100 \kms$ (Pettini, private
communication), consistent with the median velocity
width for Ly$\alpha$ of $140\kms$ measured in similar
galaxies in the HDF (Lowenthal et al. 1997).
Of course, these figures could underestimate the total velocity
dispersion, since they are dominated by emission from the central regions only.
For the present, the range of values $\sigma_v = 100$ to $320 \kms$
will be adopted, and
the sensitivity to the assumed velocity will be indicated.
In practice, this uncertainty in the velocity does
not produce an important uncertainty in the conclusions.

{\bf (3) Red radio galaxies}
\quad
An especially interesting set of objects are the reddest
optical identifications of 1-mJy radio galaxies, for which
deep absorption-line spectroscopy has proved 
that the red colours result
from a well-evolved stellar population, with a minimum
stellar age of 3.5 Gyr for 53W091 at
$z=1.55$ (Dunlop et al. 1996; Spinrad et al. 1997), and 4.0 Gyr for
53W069 at $z=1.43$ (Dunlop 1998; Dey et al. 1998). Such ages push the
formation era for these galaxies back to extremely high
redshifts, and it is of interest to ask what level of small-scale
power is needed in order to allow this early formation.

Two extremely red galaxies were found at $z=1.43$ and 1.55, 
over an area $1.68\times 10^{-3}\; \rm sr$, so a minimal
comoving density is from one galaxy in this redshift range:
\japeqn
N(\Omega=1) \gs 10^{-5.87} \; (\mpcoh)^{-3}.
\japeqn
This figure is comparable to
the density of the richest Abell clusters, and is thus in reasonable
agreement with the discovery that rich high-redshift
clusters appear to contain radio-quiet examples
of similarly red galaxies (Dickinson 1995).

Since the velocity dispersions of these galaxies are
not observed, they must be inferred indirectly. This
is possible because of the known present-day Faber-Jackson
relation for ellipticals. For 53W091, the large-aperture
absolute magnitude is
\japeqn
M_V(z=1.55\mid \Omega=1) \simeq -21.62 -5 \log_{10} h
\japeqn
(measured direct in the rest frame).
According to Solar-metallicity  spectral synthesis models, 
this would be expected to fade
by about 0.9 mag. between $z=1.55$ and the present, for
an $\Omega=1$ model of present age 14 Gyr
(note that Bender et al. 1996 have observed a shift
in the zero-point of the $M-\sigma_v$ relation out to $z=0.37$
of a consistent size).
If we compare these numbers with the $\sigma_v$ -- $M_V$
relation for Coma ($m-M=34.3$ for $h=1$) taken from
Dressler (1984), this predicts velocity dispersions in the range
\japeqn
\sigma_v= 222 \; {\rm to}\; 292 \; \kms.
\japeqn
This is a very reasonable range for a giant elliptical,
and it adopted in the following analysis.

Having established an abundance and an equivalent circular velocity
for these galaxies, the treatment of them will differ in one
critical way from the Lyman-$\alpha$ and Lyman-limit galaxies.
For these, the normal Press-Schechter approach assumes 
the systems under study to be newly born. For
the Lyman-$\alpha$ and Lyman-limit galaxies, 
this may not be a bad approximation,
since they are evolving rapidly and/or display high levels of star-formation
activity. For the radio galaxies, conversely, 
their inactivity suggests that they may have existed as
discrete systems at redshifts much higher than $z\simeq 1.5$.
The strategy will therefore be to
apply the Press-Schechter machinery at some unknown formation
redshift, and see what range of redshift gives a consistent
degree of inhomogeneity.

\japfig{2}{3}
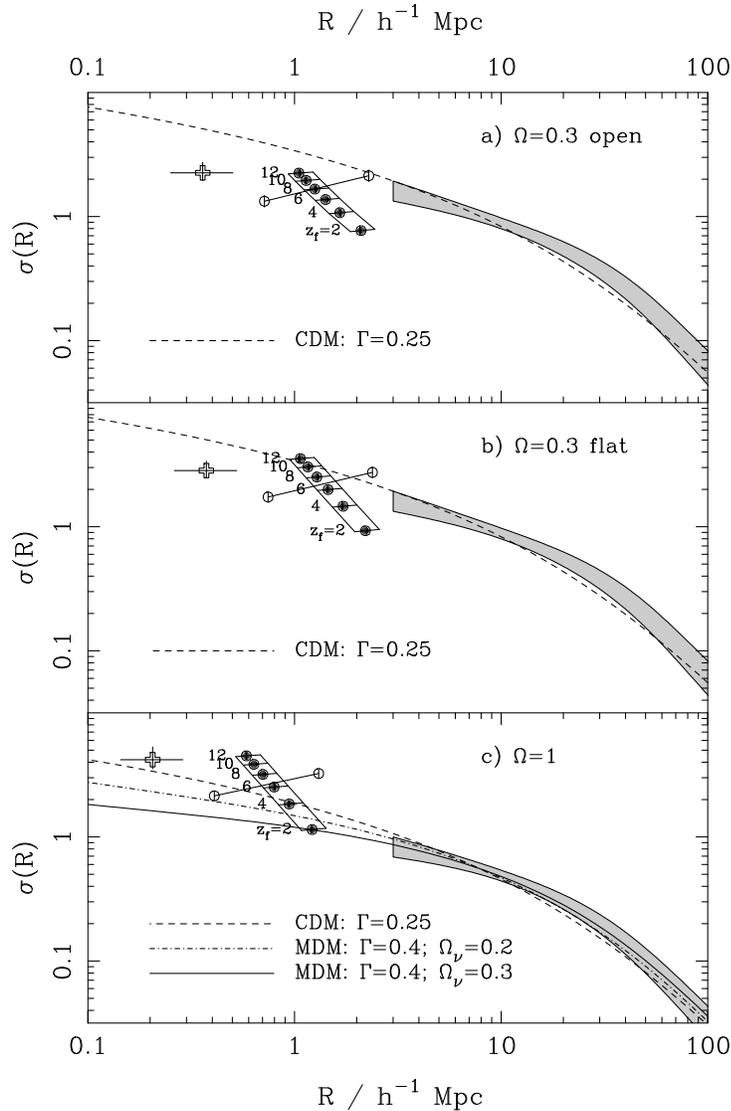
{The present-day linear 
fractional rms fluctuation in density averaged in
spheres of radius $R$. The data points are
Lyman-$\alpha$ galaxies (open cross) and
Lyman-limit galaxies (open circles)
The diagonal band with solid points shows red radio
galaxies with assumed collapse redshifts 2, 4, \dots 12.
The vertical error bars show the effect of a change in abundance by a factor 2.
The horizontal errors correspond to different choices for the
circular velocities of the dark-matter haloes that host the galaxies.
The shaded region at large $R$ gives the results inferred
from galaxy clustering.
The lines show CDM and MDM predictions,
with a large-scale normalization of $\sigma_8=0.55$ for $\Omega=1$
or $\sigma_8=1$ for the low-density models.}

\sec{The small-scale fluctuation spectrum}

\ssec{The empirical spectrum}

Fig. 2 shows the $\sigma(R)$ data which result
from the Press-Schechter analysis, for three
cosmologies. The $\sigma(R)$ numbers measured at various
high redshifts have been translated to $z=0$ using the
appropriate linear growth law for density perturbations.

The open symbols give the results for the
Lyman-limit (largest $R$) and Lyman-$\alpha$ (smallest $R$)
systems. The approximately horizontal error bars show the effect of the
quoted range of velocity dispersions for a fixed
abundance; the vertical
errors show the effect of changing the abundance by a factor 2 at fixed
velocity dispersion.
The locus implied by the red radio galaxies sits
in between. The different points show the effects of
varying collapse redshift: $z_c=2, 4, \dots, 12$
[lowest redshift gives lowest $\sigma(R)$]. Clearly, collapse
redshifts of 6 -- 8 are favoured for consistency
with the other data on high-redshift galaxies, independent
of theoretical preconceptions and independent of the
age of these galaxies. This level of power
($\sigma[R]\simeq 2$ for $R\simeq 1 \mpcoh$) is also
in very close agreement with the level of power required to
produce the observed structure in the Lyman alpha forest
(Croft et al. 1998), so there is a good case to be made that
the fluctuation spectrum has now been measured in a 
consistent fashion down to below $R\simeq 1\mpcoh$.

The shaded region at larger $R$ shows the results
deduced from clustering data (Peacock 1997). 
It is clear an $\Omega=1$ universe requires the power spectrum
at small scales to be higher than would be expected on the basis of an
extrapolation from the large-scale spectrum. Depending on assumptions
about the scale-dependence of bias, such a `feature'
in the linear spectrum  may also
be required in order to satisfy the small-scale present-day 
nonlinear galaxy clustering (Peacock 1997).
Conversely, for low-density models, the empirical small-scale
spectrum appears to match reasonably smoothly onto the large-scale data.

Fig. 2 also compares the empirical data with various physical power
spectra. A CDM model (using the transfer 
function of Bardeen et al. 1986) with shape parameter
$\Gamma=\Omega h=0.25$ is shown as a reference for all models.
This appears to have approximately the correct shape,
although it overpredicts the level of small-scale
power somewhat in the low-density cases. A better empirical
shape is given by MDM with $\Omega h\simeq 0.4$ and $\Omega_\nu\simeq 0.3$.
However, this model only makes physical sense in a universe
with high $\Omega$, and so it is only shown as
the lowest curve in Fig. 2c, reproduced from the
fitting formula of Pogosyan \& Starobinsky (1995; see also
Ma 1996). This curve fails to
supply the required small-scale power,
by about a factor 3 in $\sigma$; lowering
$\Omega_\nu$ to 0.2 still leaves a very large discrepancy.
This conclusion is in agreement with e.g. Mo \& Miralda-Escud\'e (1994),
Ma \& Bertschinger (1994), Ma et al. (1997) and Gardner et al. (1997).

All the models in Fig. 2 assume $n=1$; in fact, consistency with the COBE
results for this choice of $\sigma_8$ 
and $\Omega h$ requires a significant
tilt for flat low-density CDM models, $n\simeq 0.9$ (whereas open CDM
models require $n$ substantially above unity).
Over the range of scales probed by LSS,
changes in $n$ are largely degenerate with changes in $\Omega h$,
but the small-scale power is more sensitive to tilt
than to $\Omega h$. Tilting the $\Omega=1$ models is
not attractive, since it increases the tendency for model
predictions to lie below the data. However,
a tilted low-$\Omega$ flat CDM model would agree moderately
well with the data on all scales, with the exception of the
`bump' around $R\simeq 30 \mpcoh$. Testing the reality of this
feature will therefore be an important task for future
generations of redshift survey.

\ssec{Collapse redshifts and ages for red radio galaxies}

Are the collapse redshifts 
inferred above consistent with the
age data on the red radio galaxies? First bear in mind that
in a hierarchy some of the stars in a galaxy will inevitably
form in sub-units before the epoch of collapse.
At the time of final collapse,
the typical stellar age will be some fraction $\alpha$ of
the age of the universe at that time:
\japeqn
{\rm age} = t(z_{\rm obs}) - t(z_c) + \alpha t(z_c).
\japeqn
We can rule out $\alpha=1$ (i.e. all stars forming in small subunits  just
after the big bang). For present-day
ellipticals, the tight colour-magnitude relation
only allows an approximate doubling of the mass
through mergers since the termination of star formation
(Bower at al. 1992). This corresponds to $\alpha\simeq 0.3$
(Peacock 1991). A non-zero $\alpha$ just corresponds to
scaling the collapse redshift as
\japeqn
{\rm apparent}\ (1+z_c)\propto (1-\alpha)^{-2/3},
\japeqn
since $t\propto (1+z)^{-3/2}$
at high redshifts for all cosmologies.
For example, a galaxy which collapsed at $z=6$ would have
an apparent age corresponding to a collapse redshift of 7.9 for $\alpha=0.3$.

Converting the ages for the galaxies to an apparent collapse
redshift depends on the cosmological model, but particularly on $H_0$.
Some of this uncertainty may be circumvented by fixing the age of the
universe. After all, it is of no interest to ask about formation
redshifts in a model with e.g. $\Omega=1$, $h=0.7$ when the whole
universe then has an age of only 9.5 Gyr. If $\Omega=1$ is to be tenable
then either $h<0.5$ against all the evidence or there must be an error
in the stellar evolution timescale. If the stellar timescales
are wrong by a fixed factor, then these two possibilities
are degenerate. It therefore makes sense to measure galaxy ages
only in units of the age of the universe -- or, equivalently, 
to choose freely an apparent Hubble constant which gives the
universe an age comparable to that inferred for globular clusters.
In this spirit, Fig. 3 gives
apparent ages as a function of effective collapse redshift for
models in which the age of the universe is forced to be 14 Gyr
(e.g. Jimenez et al. 1996).

\japfig{3}{1}
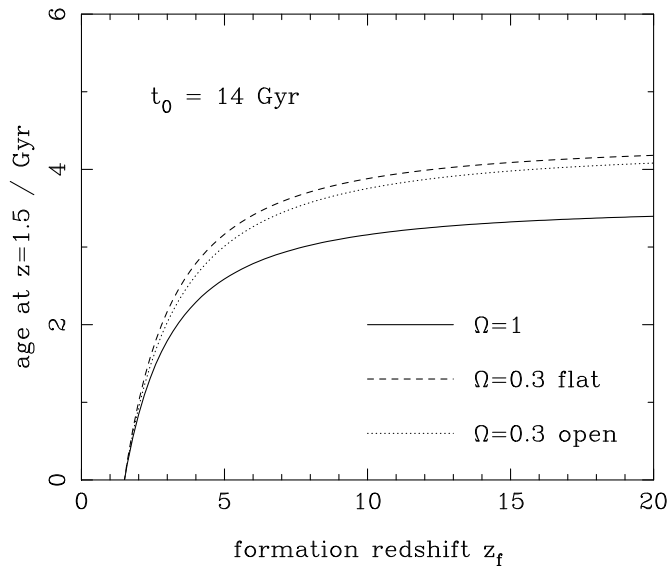
{The age of a galaxy at $z=1.5$, as a function of its
collapse redshift (assuming an instantaneous burst of star formation).
The various lines show $\Omega=1$ [solid]; open $\Omega=0.3$ [dotted];
flat $\Omega=0.3$ [dashed]. In all cases, the present
age of the universe is forced to be 14 Gyr.}

This plot shows that the ages of the red radio galaxies are
not permitted very much freedom. 
Formation redshifts in the range 6 to 8
predict an age of close to 3.0 Gyr for $\Omega=1$,
or 3.7 Gyr for low-density models, irrespective of
whether $\Lambda$ is nonzero.
The age-$z_c$ relation is rather flat, and this gives
a robust estimate of age once we have some idea of $z_c$
through the abundance arguments.
It is therefore rather satisfying that the ages inferred from
matching the rest-frame UV spectra of these galaxies
are close to the above figures.

\ssec{The global picture of galaxy formation}

It is interesting to note that it has been
possible to construct a consistent picture which
incorporates both the large numbers of star-forming
galaxies at $z\ls 3$ and the existence of old systems
which must have formed at very much larger redshifts.
A recent conclusion from the numbers of Lyman-limit
galaxies and the star-formation rates seen at $z\simeq 1$
has been that the global history of star formation
peaked at $z\simeq 2$ (Madau et al. 1996). This leaves open two possibilities
for the very old systems: either they are the rare precursors
of this process, and form unusually early, or they are
a relic of a second peak in activity at higher redshift,
such as is commonly invoked for the origin of all
spheroidal components.
While such a bimodal history of star formation
cannot be rejected, the rareness of the
red radio galaxies indicates that there is no difficulty
with the former picture. This can be demonstrated quantitatively
by integrating the total amount of star formation at high redshift.
According to Madau et al., The star-formation rate at $z=4$ is
\japeqn
\dot \rho_* \simeq 10^{7.3}h\; M_\odot\,{\rm Gyr}^{-1}\, {\rm Mpc}^{-3},
\japeqn
declining roughly as $(1+z)^{-4}$. This is probably a underestimate
by a factor of at least 3, as indicated by suggestions of dust in
the Lyman-limit galaxies (Pettini et al. 1997), and by the prediction
of Pei \& Fall (1995), based on high-$z$ element abundances.
If we scale by a factor 3, and integrate to find the total density
in stars produced at $z>6$, this yields
\japeqn
\rho_*(z_{f}>6) \simeq 10^{6.2} M_\odot\,{\rm Mpc}^{-3}.
\japeqn
Since the red mJy galaxies have a density of $10^{-5.87}h^3 {\rm Mpc}^{-3}$
and stellar masses of order $10^{11}\, M_\odot$, there is clearly no
conflict with the idea that these galaxies are 
the first stellar systems of $L^*$ size which form
en route to the general era of star and galaxy formation.

\japfig{4}{1}
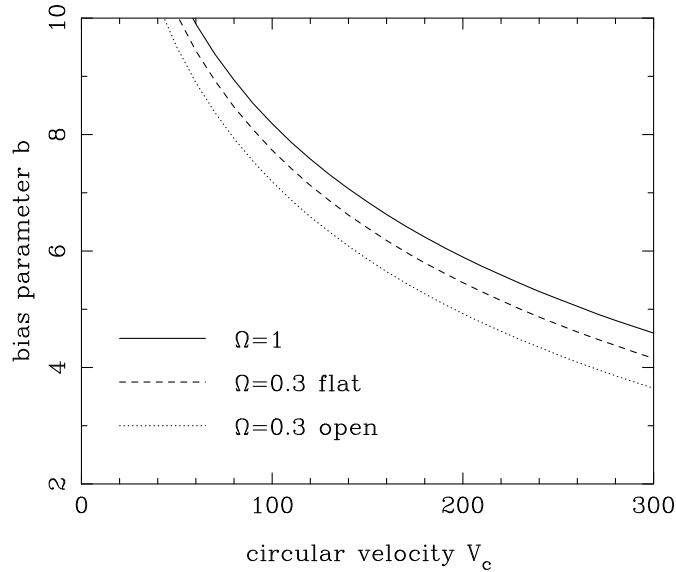
{The bias parameter at $z=3.2$ predicted for the
Lyman-limit galaxies, as a function of their assumed 
circular velocity. Dotted line shows $\Omega=0.3$ open;
dashed line is $\Omega=0.3$ flat; solid line is $\Omega=1$.
A substantial bias in the region of $b\simeq 6$ is predicted
rather robustly.}

\ssec{Predictions for biased clustering at high redshifts}

An interesting aspect of these results is that the
level of power on 1-Mpc scales is only moderate:
$\sigma(1\mpcoh)\simeq 2$. At $z\simeq 3$, the
corresponding figure would have been much lower,
making systems like the Lyman-limit galaxies rather
rare. For Gaussian fluctuations, as assumed in the
Press-Schechter analysis, such systems will be
expected to display spatial correlations which are
strongly biased with respect to the underlying mass.
The linear bias parameter depends on the rareness of
the fluctuation and the rms of the underlying field as
\japeqn
b=1+{\nu^2-1\over \nu\sigma}= 1+ {\nu^2-1\over \delta_c}
\japeqn
(Kaiser 1984; Cole \& Kaiser 1989; Mo \& White 1996),
where $\nu = \delta_c/\sigma$, and $\sigma^2$ is the
fractional mass variance at the redshift of interest.

In this analysis, $\delta_c=1.686$ is assumed. 
Variations in this number of order 10 per cent have
been suggested by authors who have studied the
fit of the Press-Schechter model to numerical data.
These changes would merely scale $b-1$ by a small amount;
the key parameter is $\nu$, which is set entirely by
the collapsed fraction. For the Lyman-limit galaxies,
typical values of this parameter are $\nu\simeq 3$,
and it is clear that very substantial values of bias
are expected, as illustrated in Fig. 4.

This diagram shows how the predicted bias parameter
varies with the assumed circular velocity, for a number density of
galaxies fixed at the level observed by Steidel et al. (1996).
The sensitivity to cosmological parameter is only
moderate; at $V_c=200\kms$, the predicted bias is $b\simeq 4.6$, 5.5, 5.8
for the open, flat and critical models respectively.
These numbers scale approximately as $V_c^{-0.4}$, and
$b$ is within 20 per cent of 6 for most plausible
parameter combinations.
Strictly, the bias values determined here are upper
limits, since the numbers of collapsed haloes of this
circular velocity could in principle greatly exceed the
numbers of observed Lyman-limit galaxies. However, the
undercounting would have to be substantial: increasing the
collapsed fraction by a factor 10 reduces the implied bias
by a factor of about 1.5. A substantial bias seems
difficult to avoid, as has been pointed out in the context of CDM
models by Baugh et al. (1998).

Comparing the bias values in Fig. 4 with those observed
directly (section 2b), we see that
the observed value of $b$ is quite close to the prediction
in the case of $\Omega=1$
-- suggesting that the simplest interpretation of these
systems as collapsed rare peaks may well be roughly correct.
Indeed, for high circular velocities
there is a danger of exceeding the predictions, and it would
create something of a difficulty for high-density models if
a velocity as high as $V_c\simeq 300 \kms$ were to be established as
typical of the Lyman-limit galaxies.
For low $\Omega$, the `observed' bias falls faster than the
predictions, so there is less danger of conflict. For a circular
velocity of $200 \kms$, we would need to say that the collapsed fraction
was underestimated by roughly a factor 10
(i.e. increase the values of $\sigma$ in Fig. 2 by a factor
of about 1.5) in order to lower the
predicted bias sufficiently, either by postulating that the
conversion from velocity to $R$ is systematically in error, or
by suggesting that there may be many haloes which are not detected
by the Lyman-limit search technique. It is hard to argue that
either of these possibilities are completely ruled out.
Nevertheless, we have reached the paradoxical conclusion that
the observed large-amplitude clustering at $z=3$ is more naturally
understood in an $\Omega=1$ model, whereas one might have expected the
opposite conclusion.

\sec{Empirical predictions for CMB anisotropies}

The recurring theme of this paper has been that it is
now possible to measure the fluctuation spectrum
empirically to an interesting precision. On large scales,
this is possible using galaxy clustering to give the shape
of the spectrum, with the cluster abundance giving the normalization.
On small scales, we have seen how information on high-redshift
galaxies gives answers that are reasonably consistent with
extrapolation of the large-scale results. 
This situation is to be contrasted with the normal
approach to measurements of CMB anisotropies, where the
results are fitted by variants on CDM models, adjusting
the parameters $(\Omega_m, \Omega_v, \Omega_{\ss B}, h, n)$.
If CMB data alone are considered, many combinations of these parameters
can fit existing results; however, in many cases the
predicted $z=0$ matter fluctuation spectrum will be
in gross disagreement with observation.

\japfig{5}{1}
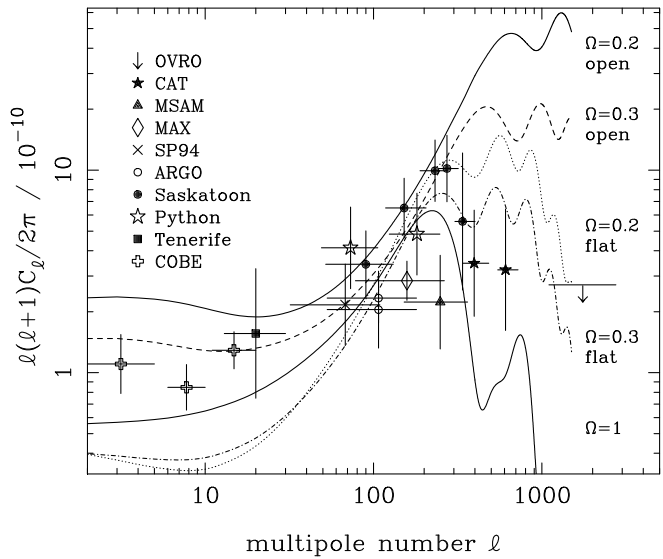
{The angular power spectrum for the temperature fluctuations
in the CMB. These predictions fix the matter power spectrum
today to have the shape inferred from galaxy clustering and
the normalization inferred from the abundance of rich clusters.
The `observed' values are adopted for the other cosmological
parameters: $h=0.65$, $\Omega_{\ss B}/\Omega=0.1$. At long
wavelengths, where no galaxy clustering data exist, the spectrum
is assumed to be scale invariant; failure to match COBE thus
indicates that tilt is required. However, the power at
$\ell\sim 2000$ is nearly spectrum independent, since this is
where the normalization scale sits. The rejection of open
models is thus very nearly model independent.}

This problem is often tackled by requiring acceptable models
to fit some statistic such as $\sigma_8$. However, an
alternative route is to recall that the CMB calculations
are entirely linear, and that they are based on the
evolution of a given Fourier mode from last scattering
at $z\simeq 1100$ to the present. The equations involved
are time-symmetric, so there is no reason why the integration
cannot be carried out backwards. If we believe that the
amplitude of gravitational potential fluctuations at
$z=0$ has been measured as a function of scale, then it
makes sense to place these fluctuations at last scattering
and deduce an empirical prediction of the CMB fluctuations.
In practice, this can be achieved by a process which
resembles `designer inflation': assume a suitable
fluctuation spectrum at $z>1100$ such that any features
in the transfer function are cancelled, leaving the
desired power spectrum at $z=0$. Described in this way,
the process sounds unnatural; however, the standard lore
suggests that perturbations are generated at $z\sim 10^{28}$,
so there is much more room at $z>1100$ for unknown extra physics
than there is at $z<1100$.

This approach still leaves free the global cosmological
parameters. The CMB results clearly depend on $\Omega_m$
and $\Omega_v$, since the inferred fluctuation spectrum
depends on these (although only weakly on $\Omega_v$).
The other parameters can be fixed at their empirical
values, taken here to be 
$h=0.65$ and $\Omega_{\ss B}/\Omega=0.1$.
For an extreme  empirical approach, no power would be
assumed beyond the largest scale at which clustering is
observed in the galaxy distribution ($k\simeq 0.02 \hompc$).
A reasonable alternative, adopted here, is to allow
the spectrum to vary with some power-law index $n$ 
on larger scales. Finally, the collisionless dark matter
is taken to be cold and the fluctuations are assumed to
be isentropic; variations of either of these assumptions would lead
to larger fluctuations.

The results of this calculation are shown in Fig. 5, which
was generated using a modified form of the CMBFAST code of
Seljak \& Zaldarriaga (1996). This looks quite different
from the plots usually seen in this field, for the
following reasons:

(1) The normalization comes direct from the
power spectrum; no attempt has been made to fit the CMB data.

(2) The models are thus not COBE normalized, although
they could be made to fit COBE by adjusting the large-wavelength
index $n$. Open models would require $n>1$, flat models $n<1$.

(3) Adjusting $n$ in this way only affects $C_\ell$
for $\ell \ls 300$, since larger multipoles project to
parts of $k$ space probed by galaxy clustering.

The last point is especially important, since it means that
it is the high-$\ell$ clustering where robust predictions
can be made. The $k\simeq 0.2 \hompc$ waves that
determine $\sigma_8$ project to $\ell\simeq 1200$ for
$\Omega=1$, or to $\ell\simeq 1200/\Omega$ and
$1200/\Omega^{0.4}$ for open and flat models respectively.
This difference in angular-diameter distance is one
half of the reason why the predictions for open models
in Fig. 5 are so much higher than the predictions for
flat models with the same parameters. The other difference
is the difference in linear growth suppression factors, which
amounts roughly to a factor $\Omega^{0.4}$ (equation 1.1).
Since the present-day power spectrum and its normalization
is highly insensitive to $\Lambda$, there is thus a very
simple recipe for predicting the CMB anisotropies for a given
open model: calculate the corresponding flat case,
boost the results  by a factor $\Omega^{-0.8}$ and
translate the spectrum to higher $\ell$ by  a factor $\Omega^{-0.6}$.
This recipe fails for the lowest multipoles, where spatial
curvature is important. However, the critical $\ell\sim 1000$
results follow this scaling almost exactly. 

The conclusion
is therefore that, for any flat model which roughly fits
the CMB data, its open counterpart will be grossly in
error, and this is just what is seen in practice.
Flat models with $\Omega\gs 0.3$ are acceptable, but open models
are qualitatively wrong unless $\Omega \gs 0.5$.
It is interesting to note that this conclusion
comes not so much from the modern data at $\ell\simeq 200$,
but from the long-standing OVRO upper limit at
$\ell\simeq 2000$. The inconsistency of this result with
most open models was noted by Bond \& Efstathiou (1984), and
all that has changed since then is that we now prefer to
use the cluster normalization, rather than the unbiased
normalization chosen by Bond \& Efstathiou. This raises
the amplitude for low-$\Omega$ models, making it that
much harder for them to get anywhere near to the data.

\sec{Conclusions}

The data on the abundances and clustering of both
radio-loud and radio-quiet galaxies at high redshift
appear to be in good quantitative agreement with
the expectation of models in which structure formation
proceeds through hierarchical merging of haloes of dark matter.
Furthermore, the existing
data yield an empirical measurement of the fluctuation spectrum
on sub-Mpc scales.
In general, this small-scale spectrum is close to
what would be expected from an extrapolation of
LSS measurements, but there are deviations in detail:
$\Omega=1$ places the small-scale data somewhat above
the LSS extrapolation, whereas open low-$\Omega$ models suffer from the
opposite problem; low-$\Omega$ $\Lambda$-dominated models
fare somewhat better, especially with a slight tilt. These last models also
account well for the $\ell\sim 1000$ CMB anisotropies
if the dark matter is assumed to be pure
CDM, normalized to COBE (whereas open models fail badly). 
Until recently, it appeared that geometrical tests
such as the supernova Hubble diagram (Perlmutter et al. 1997)
or gravitational lensing (Carroll, Press \& Turner 1992; Kochanek 1996)
were strongly inconsistent with $\Lambda$-dominated
models, so the overall situation was badly confused.
However, with recent developments in these areas now appearing
to favour a nonzero $\Lambda$ (Garnavich  et al. 1998; Chiba \& Yoshii 1997), 
it is possible that a consistent picture may be emerging.

The main remaining difficulty for $\Lambda$CDM lies in the
shape of the large-scale power spectrum measured from the APM survey
around $k=0.1 \hompc$. This is a region of the spectrum which
is well within the capability of 2dF and Sloan, so we can
confidently expect this problem to be either confirmed
or removed within the next few years.
The subject of structure formation thus stands at a critical point:
either we are close to having a `standard model' for
galaxy formation and clustering, or we may
have to accept that radical new ideas are needed.
At the current rate of observational progress, the
verdict should not be very far away.

\section*{Acknowledgements}
This paper draws on unpublished collaborative work
James Dunlop, Raul Jimenez, Ian Waddington, Hy Spinrad,
Daniel Stern, Arjun Dey \& Rogier Windhorst.
The work on CMB anisotropies was performed during a visit
to Caltech, for which thanks are due to Tony Readhead.

\section*{References}

\begin{thedemobiblio}{}
\japref Adelberger K., Steidel C., Giavalisco M., Dickinson M., Pettini M., Kellogg M., 1998, astro-ph/9804236
\japref Andreani P., Cristiani S., 1992, A\&A, 398, L13
\japref Bardeen J.M., Bond J.R., Kaiser N., Szalay A.S., 1986, {\apj}, {304}, 15 
\japref Baugh C.M., Efstathiou G., 1993, MNRAS, 265, 145
\japref Baugh C.M., Efstathiou G., 1994, MNRAS, 267, 323
\japref Baugh C.M., Cole S., Frenk C.S., Lacey C.G., 1998, ApJ, 489, 504
\japref Bender R., Ziegler B., Bruzual G., 1996, \apj, 463, L51
\japref Bond J.R., Efstathiou G., {1984}, {\apj}, {285}, {L45}
\japref Bower R.G., Lucey J.R., Ellis R.S., 1992, {MNRAS}, {254}, 601
\japref Broadhurst T.J., Taylor A.N., Peacock J.A., 1995, ApJ, 438, 49
\japref Carlberg R.G., Yee H.K.C., Ellingson E., Abraham R., Gravel P., Morris S., Pritchet C.J., 1996, ApJ, 462, 32
\japref Carlberg R.G., Cowie L.L., Songaila A., Hu E.M., 1997, ApJ, 484, 538
\japref Carroll S.M., Press W.H., Turner E.L., {1992}, {ARAA}, {30}, {499}
\japref Chiba M., Yoshii Y., {1997}, {\apj}, {490}, {L73}
\japref Clowe D., Luppino G.A., Kaiser N., Henry J.P., Gioia I.M., 1998, ApJ, 497, L61
\japref Cole S., Kaiser N., 1989, MNRAS, 237, 1127
\japref Couch W.J., Jurcevic J.S., Boyle B.J., 1993, MNRAS, 260, 241
\japref Cress C.M., Helfand D.J., Becker R.H., Gregg M.D., White R.L., 1996, ApJ, 473, 7
\japref Croft R.A.C. et al., 1998, ApJ, 495, 44
\japref Croom S.M., Shanks T., 1996, MNRAS, 281, 893
\japref Dey A., et al., 1998,  for submission to ApJ
\japref Dickinson M., 1995, in ``Fresh Views of Elliptical Galaxies'', ASP conf. ser. Vol 86,  eds A. Buzzoni, A. Renzini, A. Serrano, p283
\japref Dressler A., 1984, \apj, 281, 512
\japref Dunlop J.S., Taylor G.L., Hughes D.H., Robson E.I., 1993, MNRAS, 264, 455
\japref Dunlop J.S., Peacock J.A., Spinrad H., Dey A., Jimenez R., Stern D., Windhorst R.A., 1996, Nat, 381, 581
\japref Dunlop J.S., 1998, astro-ph/9801114
\japref Efstathiou G., 1992, {\mn}, {256}, 43P
\japref Efstathiou G., Rees M.J., 1988, {\mn}, {230}, 5P
\japref Efstathiou G., Bernstein G., Katz N., Tyson T., Guhathakurta P., 1991, ApJ, 380, 47
\japref Eke V.R., Cole S., Frenk C.S., {1996}, {MNRAS}, {282}, {263}
\japref Gardner J.P., Katz N., Weinberg D.H., Hernquist L., 1997, ApJ, 486, 42
\japref Garnavich P.M. et al., 1998, ApJ, 493, L53
\japref Haehnelt M.G., Steinmetz M., Rauch M., 1998, ApJ, 495, 647
\japref Henry J.P., Gioia I.M., Maccacaro T., Morris S., Stocke J.T., Wolter A., 1992, ApJ, 386, 408
\japref Henry J.P., 1997, ApJ, 489, L1
\japref Jimenez R., Thejl P., J\o rgensen U.G., MacDonald J., Pagel B., 1996, \mn, 282, 926
\japref Kaiser N., 1984, ApJ, 284, L9
\japref Kaiser N., 1992, ApJ, 388,272
\japref Kauffmann G., Charlot S., 1994, ApJ, 430, L97
\japref Klypin A. et al., 1995, ApJ, 444, 1
\japref Kochanek C.S., {1996}, {\apj}, {466}, {638}
\japref Lahav O., Lilje P.B., Primack J.R., Rees M.J., {1991}, {MNRAS}, {251}, {128}
\japref Lanzetta K., Wolfe A.M., Turnshek D.A., Lu L., McMahon R.G., Hazard C., 1991, {\apjs}, {77}, 1
\japref Le Brun V., Bergeron J., Boisse P., De Harveng J.M., 1996, A\&A, 321, 733
\japref Le F\`evre O., et al., 1996. ApJ, 461, 534
\japref Lilly S.J., Le F\`evre O., Hammer F., Crampton D., 1996, ApJ, 460, L1
\japref Loan A.J., Lahav O., Wall J.V., 1997, MNRAS, 286, 994
\japref Lowenthal J.D., et al., 1997, {\apj}, 481, 673
\japref Luppino G.A., Gioia I.M., 1995, ApJ, 445, L77
\japref Ma C., Bertschinger E., 1994, {\apj}, {434}, L5
\japref Ma C., 1996, \apj, 471, 13
\japref Ma C., Bertschinger E., Hernquist L., Weinberg D., Katz N., 1997, ApJ, 484, L1
\japref Madau P. et al., 1996, MNRAS, 283, 1388
\japref Maddox S. Efstathiou G., Sutherland W.J., 1996, MNRAS, 283, 1227
\japref Mo H.J., Miralda-Escud\'{e} J., 1994, {\apj}, {430}, L25
\japref Mo H.J., Fukugita M., 1996, {\apj}, {467}, L9
\japref Mo H.J., White S.D.M., 1996, MNRAS, 282, 1096
\japref Neuschaefer L.W., Windhorst R.A., Dressler A., 1991, ApJ, 382, 32
\japref Peacock J.A., 1991, in ``Physical Cosmology'', proc. 2$^{nd}$ Rencontre de Blois, eds A. Blanchard, L. Celnekier, M. Lachi\`eze-Rey \& J. Tr\^an Thanh V\^an (Editions Fronti\`eres), p337
\japref Peacock J.A., 1997, \mn, 284, 885
\japref Peacock J.A., Jimenez R., Dunlop J.S., Waddington I., Spinrad H., Stern D., Dey A., Windhorst R.A., 1998, astro-ph/9801184
\japref Peebles P.J.E., 1980, The Large-Scale Structure of the Universe.  Princeton Univ. Press, Princeton, NJ
\japref Pei Y.C., Fall S.M., 1995, ApJ, 454, 69
\japref Pettini M., Steidel C.C., Dickinson M., Kellogg, M., Giavalisco M., Adelberger K.L., 1997, astro-ph/9707200
\japref Perlmutter S. {et al.}, {1997b}, {\apj}, {483}, {565}
\japref Pogosyan D.Y., Starobinsky A.A., 1995, \apj, 447, 465
\japref Press W.H., Schechter P., 1974, {\apj}, {187}, 425
\japref Roche N., Shanks T., Metcalfe N., Fong R., 1993, MNRAS, 263, 360
\japref Saunders W., Rowan-Robinson M., Lawrence A., {1992}, {MNRAS}, {258}, {134}
\japref Seljak U., Zaldarriaga M., {1996}, {\apj}, {469}, {437}
\japref Shanks T., Boyle B.J., 1994, MNRAS, 271, 753
\japref Spinrad H., Dey A., Stern D., Dunlop J., Peacock J., Jimenez R., Windhorst R., 1997, ApJ, 484, 581
\japref Steidel C.C., Giavalisco M., Pettini M., Dickinson M., Adelberger K.L., 1996, \apj, 462, L17
\japref Steidel C.C., Adelberger K.L., Dickinson M., Giavalisco M., Pettini M., Kellogg M., 1998, ApJ, 492, 428
\japref Stephens  A.W., Schneider D.P., Schmidt M., Gunn J.E., Weinberg D.H., 1997, AJ, 114, 41
\japref Storrie-Lombardi L.J., McMahon R.G., Irwin M.J., 1996, MNRAS, 283, L79
\japref Taylor G.L., Dunlop J.S., Hughes D.H., Robson E.I., 1996, MNRAS, 283, 930
\japref Tytler D., Fan X.-M., Burles S., 1996, Nat, 381, 207
\japref Viana P.T., Liddle A.R., 1996, MNRAS, 281, 323
\japref Walker T.P., Steigman G., Schramm D.N., Olive K.A., Kang H.S., 1991, {\apj}, {376}, 51
\japref White S.D.M., Efstathiou G., Frenk C.S., 1993, {\mn}, {262}, 1023
\end{thedemobiblio}{}
\end{document}